\documentclass[a4paper]{amsart}
\usepackage{amssymb}
\usepackage{amsmath}
\usepackage{amsthm}
\usepackage{amsfonts}
\usepackage{graphicx}
\usepackage{float}

\newtheorem{theorem}{Theorem}

\newtheorem{conjecture}{Conjecture}

\newtheorem{example}[theorem]{Example}

\numberwithin{equation}{section}

\begin{document}

\title[Is the min  value of an option on variance generated by local
vol?]{Is the minimum value of an option on variance generated by local
volatility?}
\author{Mathias Beiglb\"ock}
\author{Peter Friz}  
\author{Stephan Sturm}

\address{Mathias Beiglb\"ock\endgraf Fakult\"at f\"ur Mathematik, Universit\"at Wien\endgraf
Nordbergstra\ss e 15\\ 1090 Wien, Austria}
\email{mathias.beiglboeck@univie.ac.at}
\address{Peter Friz \endgraf Institut f\"ur Mathematik, TU Berlin\endgraf
Stra\ss e des 17. Juni 136 \\ 10623 Berlin, Germany\endgraf
{\it and} Weierstra\ss --Institut f\"ur Angewandte Analysis und Stochastik
\endgraf
Mohrenstra\ss e 39, 10117 Berlin, Germany }
\email{friz@math.tu-berlin.de {\it and} friz@wias-berlin.de}
\address{Stephan Sturm\endgraf Department of Operations Research and Financial Engineering\endgraf Princeton University\endgraf
116 Sherrerd Hall\\ Princeton, NJ 08544 }
\email{ssturm@princeton.edu}

\thanks{The first author acknowledges support from the Austrian Science
Fund (FWF) under grant P21209. The second and the third author (affiliated to TU Berlin while this work was started) acknowledge support by MATHEON. All authors thank Gerard Brunick, Johannes Muhle-Karbe and Walter Schachermayer for useful comments.}
\maketitle

\begin{abstract}
We discuss the possibility of obtaining model-free bounds on volatility
derivatives, given present market data in the form of a calibrated local
volatility model. A counter-example to a wide-spread conjecture is given. 

\medskip

 \emph{keywords: local vol, Dupire's formula; MSC: 91G99; JEL: G10. }

\end{abstract}

\section{Introduction}

``... \textit{it has been conjectured that the minimum possible value of an
option on variance is the one generated from a local volatility model fitted
to the volatility surface.}''; Gatheral \cite[page 155]{Gath06}.

Leaving precise definitions to below, let us clarify that an 
\textit{option on variance} refers to a derivative whose payoff is a convex
function $f$ of \textit{total realized variance}. Turning from convex to concave, this
conjecture, if true, would also imply that that the maximum possible
value of a \textit{volatility swap} ($f(x)=x^{1/2}$) is the one generated from a local volatility
model fitted to the volatility surface. Given the well-documented
model-risk in pricing volatility swaps, such bounds are of immediate practical interest.

The mathematics of local volatility theory (\`{a} la Dupire, Derman, Kani, ...) is intimately related to the following

\begin{theorem}[\protect\cite{Gyon86}]
\label{ThmGy}Assume $dY_{t}=\mu \left( t,\omega \right) dt+\sigma \left(
t,\omega \right) dB_t$ is a multi-dimensional It\^{o}-process where $B$ is a multi-dimensional Brownian motion, $\mu
,\sigma $ are progressively measurable, bounded and $\sigma \sigma ^{T}\geq
\varepsilon ^{2}I$ for some $\varepsilon >0$ ($\sigma^T$ denotes the transpose of $\sigma$). Then%
\begin{equation}\label{GyongyT}
d\tilde{Y_t}=\mu _{loc}\left( t,\tilde{Y_t}\right) dt+\sigma _{loc}\left( t,%
\tilde{Y_t}\right) d\tilde{B}_t, \qquad \tilde{Y}_0 = Y_0,
\end{equation}%
\begin{eqnarray*}
 \mu _{loc}\left( t,y\right) &=&E\left[ \mu \left( t,\omega \right) |Y_{t}=y%
\right], \\
\sigma _{loc}\left( t,y\right) &=&E\left[ \sigma \left( t,\omega
\right) \sigma^T \left( t,\omega
\right) |Y_{t}=y\right]^\frac{1}{2},
\end{eqnarray*}%
(where the power $\frac{1}{2}$ denotes the positive square root of a positive definite matrix) has a weak solution $\tilde Y_t$ such that $\tilde{Y}_{t}\overset{law}{=}Y_{t}$ for all fixed $t$.
\end{theorem}
We will apply theorem \ref{ThmGy} only in the simple one dimensional (resp.\ two dimensional in section 4 below) setting where it is well known that the solution to \eqref{GyongyT} is unique (cf. \cite{Kryl69} or \cite[Capter 7]{StVa06}).

A generic stochastic volatility model (already written under the appropriate
equivalent martingale measure and with suitable choice of num\'{e}raire) is
of the form $dS=S\sigma dB$ where $\sigma =\sigma \left( t,\omega \right) $
is the (progressively measurable) instantaneous volatility process. (It will
suffice for our application to assume $\sigma $ to be bounded from above and
below by positive constants.) Arguing on log-price $X=\log S$ rather than $S
$, 
\begin{equation}
dX_{t}=\sigma \left( t,\omega \right) dB_{t}-\left( \sigma ^{2}\left(
t,\omega \right) /2\right) dt,  \label{StochVolModelX}
\end{equation}%
a classical application of theorem \ref{ThmGy} yields the following
Markovian projection result\footnote{%
Let us quickly remark that Markovian projection techniques have led recently
to a number of new applications (see \cite{Pit06}, for instance).}: the
(weak) solution to 
\begin{eqnarray} \label{VolProjection}
d\tilde{X}_{t} &=&\sigma _{loc}\left( t,\tilde{X}\right) d\tilde{B}_{t}-\left(
\sigma _{loc}^{2}\left( t,\tilde{X}_{t}\right) /2\right) dt, \qquad \tilde{X}_0 = X_0,\\
\notag ( \sigma _{loc}\left( t,x\right)  &=&E\left[ \sigma^2
\left( t,\omega \right) |X_{t}=x\right]^\frac{1}{2} ) 
\end{eqnarray}%
has the one-dimensional marginals of the original process $X_{t}$.
Equivalently\footnote{%
The abuse of notation, by writing both $\sigma _{loc}( t,\tilde{X}%
) $ and $\sigma _{loc}( t,\tilde{S}) $, will not cause
confusion.}, the process $\tilde{S}=\exp \tilde{X},$ 
\begin{equation*}
d{\tilde{S}}_{t}={\sigma }_{loc}(t,{\tilde{S}}_{t}){\tilde{S}}%
_{t}\,dB_{t},
\end{equation*}%
known as (Dupire's)\textit{\ local volatility model}, gives rise to
identical prices of all European call options $C\left( T,K\right)$.\footnote{We emphasize that $C(T,K)$ denotes the price at time $t=0$  of European call with maturity $T$ and strike $K$.} It easily follows
that ${\sigma }_{loc}^{2}(t,{\tilde{S}})$ is given by \textit{Dupire's formula}%
\begin{equation}
{\sigma }_{loc}^{2}(T,{\tilde{S}})|_{\tilde{S}=K}=2\frac{\partial _{T}C}{%
K^{2}\partial _{KK}C}.  \label{DupiresFormula}
\end{equation}

\textit{Volatility derivatives} are options on realized variance; that is,
the payoff is given by some function $f$ of realized variance. The latter is
given by 
\begin{equation*}
V_{T}:=\langle \log S\rangle _{T}=\left\langle X\right\rangle
_{T}=\int_{0}^{T}\sigma ^{2}(t,\omega )\,dt;
\end{equation*}%
in the model $dS=\sigma \left( t,\omega \right) S\, dB$ and by%
\begin{equation*}
\tilde{V}_{T}:=\langle \log \tilde{S}\rangle _{T}=\langle \tilde{X}%
\rangle _{T}=\int_{0}^{T}\sigma _{loc}^{2}(t,\tilde{X}_{t})\,dt
\end{equation*}%
in the corresponding local volatility model.

Common choices of $f$ are $f\left( x\right) =x$, the \textit{variance swap}, 
$f\left( x\right) =x^{1/2}$, the volatility swap, or simply $f\left(
x\right) =(x-K)^{+}$, a \textit{call-option on realized variance}. See \cite%
{FG05} for instance. As is well-known, see e.g.\ \cite{Gath06}, the pricing
of a variance swap, assuming continuous dynamics of $S$ such as those
specified above, is model free in the sense that it can be priced in terms
of a log-contract; that is, a European option with payoff $\log S_{T}$. In
particular, it follows that%
\begin{equation*}
E\left[ \tilde{V}_{T}\right] =E\left[ V_{T}\right] .
\end{equation*}%
Of course this can also be seen from (\ref{VolProjection}), after exchanging 
${E}$ and integration over $\left[ 0,T\right] $. Passing from $V_{T}$ to $f(V_{T})$ for general $f$ this
is not true, and the resulting differences are known in the industry as 
\textit{convexity adjustment. } We can now formalize the conjecture given in the first lines of the introduction\footnote{%
It is tacitly assumed that $f\left( V_{T}\right) ,f( \tilde{V}%
_{T}) $ are integrable.}.

\begin{conjecture}
\label{ConjDup}For any convex $f$ one has ${E}\big[
f\big( {\tilde V_{T}}\big) \big] \leq {E}\big[ f\big( {V_{T}}\big) %
\big].$
\end{conjecture}

Our contribution is twofold: first we discuss a simple (toy) example which
provides a counterexample to the above conjecture; secondly we refine our
example using a 2-dimensional Markovian projection (which may be interesting
in its own right) and thus construct a perfectly sensible Markovian
stochastic volatility model in which the conjectured result fails. All this
narrows the class of possible dynamics for $S$ for which the conjecture can
hold true and so should be a useful step towards positive answers.

\section{Idea and numerical evidence}

\begin{example}\label{toyexample}
Consider a Black--Scholes ``mixing'' model $dS=S\sigma dB, S_0=1$ with time horizon $T=3$ in which $\sigma ^{2}\left( t,\omega \right) $ is given by $\sigma
_{+}^{2}(t)$ or $\sigma _{-}^{2}(t),$ 
\begin{equation*}
\sigma _{+}^{2}(t):=%
\begin{cases}
2 & \mbox{if $t\in [0,1]$,} \\ 
3 & \mbox{if $t\in ]1,2]$,} \\ 
1 & \mbox{if $t\in ]2,3]$,}%
\end{cases}%
\quad \sigma _{-}^{2}(t):=%
\begin{cases}
2 & \mbox{if $t\in [0,1]$,} \\ 
1 & \mbox{if $t\in ]1,2]$,} \\ 
3 & \mbox{if $t\in ]2,3]$,}%
\end{cases}%
\end{equation*}%
depending on a fair coin flip $\epsilon=\pm 1$ (independent of $B$). 
Obviously $V=V_3=\int_0^3 \sigma^2\, dt\equiv 6$ in this example, hence $E\big[ 
(V-6)^+\big] = (V-6)^+=0$. On the other
hand, the local volatility is explicitly computable (cf.\ the following
section) and one can see from simple Monte Carlo simulations that for $\tilde V= \tilde V_3$%
\begin{equation*}
E\left[(\tilde V-6)^+\right] \approx0.026 >0 
\end{equation*}%
thereby (numerically) contradicting conjecture \ref{ConjDup}, with $f\left( x\right) =(x-6)^+$. 
\end{example}

Our analysis of this toy model is simple enough: in section \ref{TheAnalysis} below we prove that $P%
[ \tilde{V}=6] \neq 1$. Since $E [\tilde{V}]=E [{V}]=6$ and $(x-6)^+$ is strictly convex at $x=6$, Jensen's inequality then tells us that $E[ (\tilde V-6)^+] >0=E\left[ (V-6)^+\right] $. 

\medskip

The reader may note that an even simpler construction would be possible, i.e.\ one could simply leave out the interval $[0,1]$ where $\sigma_+^2$ and $\sigma_-^2$ coincide. We decided not to do so for two reasons. First, insisting on $\sigma_+^2(t)=\sigma_-^2(t)$ for $t\in [0,1]$ leads to well behaved coefficients of the SDE describing the local volatility model. Second, we will use the present setup to obtain a complete model contradicting conjecture \ref{ConjDup} at the end of the next section.

\section{Analysis of the toy example}\label{TheAnalysis}
We  recall that it suffices to show that $\tilde{V}=\int_{0}^{3}{\sigma }_{loc}^{2}(t,{\tilde X}_{t})\,dt$ is not a.s.\ equal to $V\equiv 6$. The distribution of $X_{t}$ is
simply the mixture of two normal distributions. More explicitly, $X_t=I_{\{\epsilon=+1\}} X_{t,+}+I_{\{\epsilon=-1\}} X_{t,-}$, 
$$X_{t,\pm}=\int_0^t \sigma_{\pm}(s) \, dB_s - \tfrac12\int_{0}^{t}\sigma _{\pm }^{2}(s)\,ds\ \ \sim\ \ N\Big(\tfrac12\Sigma _{\pm }(t),\Sigma _{\pm }(t)\Big),$$ 
where $\Sigma _{\pm }(t):=\int_{0}^{t}\sigma _{\pm }^{2}(s)\,ds$. Thus  ${\sigma }_{loc}^{2}(t,x)=%
{E}[\sigma ^{2}(t,\omega )|X_{t}=x]$ is given by\footnote{More general expression for local volatility are found in \cite[Chapter 4]{BrMe06} and \cite{Lee01, HL09}. Note the necessity to keep $\sigma^2(.,\omega)$ constant on some interval $[0,\varepsilon$], for otherwise the local volatility surface is not Lipschitz in $x$, uniformly as $t \to 0$.
} 
\begin{equation}\label{ExplicitSigma}
{\sigma }_{loc}^{2}(t,x)=\frac{\tfrac{\sigma _{+}^{2}(t)}{\sqrt{\Sigma
_{+}(t)}}\exp \big[-\tfrac{(x+\Sigma _{+}(t)/2)^{2}}{2\Sigma _{+}(t)}\big]+%
\tfrac{\sigma _{-}^{2}(t)}{\sqrt{\Sigma _{-}(t)}}\exp \big[-\tfrac{(x+\Sigma
_{-}(t)/2)^{2}}{2\Sigma _{-}(t)}\big]}{\tfrac{1}{\sqrt{\Sigma _{+}(t)}}\exp %
\big[-\tfrac{(x+\Sigma _{+}(t)/2)^{2}}{2\Sigma _{+}(t)}\big]+\tfrac{1}{\sqrt{%
\Sigma _{-}(t)}}\exp \big[-\tfrac{(x+\Sigma _{-}(t)/2)^{2}}{2\Sigma _{-}(t)}%
\big]}.
\end{equation}%
Since ${{\sigma_{loc}}}={{\sigma_{loc}}}(s,x)$ is bounded, measurable in $t$ and Lipschitz in $x$ (uniformly w.r.t.\  $t$) and
bounded away from zero it follows from \cite[Theorem 5.1.1]{StVa06} that the
SDE 
\begin{equation*}
d\tilde{X}_{t}=\sigma _{loc}\left( t,\tilde{X}\right) dB_{t}-\tfrac12 \sigma
_{loc}^{2}\left( t,\tilde{X}_{t}\right)\, dt
\end{equation*}%
has a unique strong solution\ (started from $\tilde{X}_{0}=0$, say). Since ${{\sigma_{loc}}}$ is uniformly bounded away from $0$ it follows that the
process $(\tilde{X}_{t})$ has \emph{full support}, i.e.\ for every
continuous $\varphi :[0,3]\rightarrow \mathbb{R},\,\varphi (0)=0$
and every $\varepsilon >0$  
\begin{equation*}
P[\Vert \tilde{X}_{t}-\varphi (t)\Vert _{\infty ;[0,T]}\leq \varepsilon ]>0.
\end{equation*}%
Indeed, there are various ways to see this: one can apply Stroock--Varadhan's support theorem, in the form of \cite[Theorem 6.3]{Pins95} (several simplifications arise in the proof thanks to the one-dimensionality of the present problem); alternatively, one can employ localized lower heat kernel bounds (\`{a} la Fabes--Stroock \cite{FaSt86}) or exploit that the It\^{o}-map is continuous here (thanks to Doss--Sussman, see for instance \cite[page 180]{RoWi00}) and deduce the support statement from the full support of $B$.

\begin{figure}[tbh]
\centering
\includegraphics[height=7cm, angle=0, origin=c]{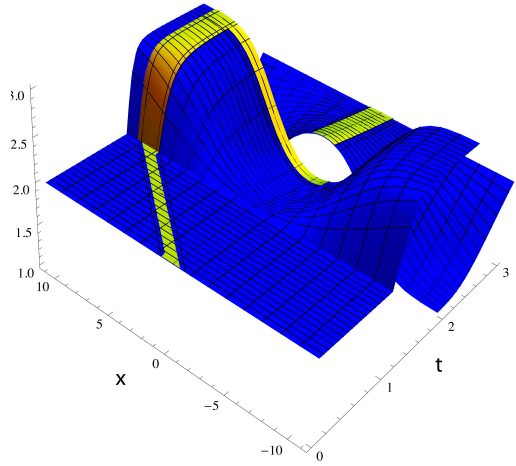}
\caption{Time evolution of local variance $\sigma_{loc}^{2}(t,x)$ in dependence of log-moneyness. The bright strip indicates a set of paths with realized variance strictly larger than $6$.}
\label{fig:localvol4.eps}
\end{figure}

Figure 1 illustrates the dependence of $\sigma_{loc}^{2}(t,x)$ on time $t$
and log-moneyness $x$. To gain our end of proving that $\tilde{V}\left(
\omega \right) =\int_{0}^{3}{\sigma }_{loc}^{2}(t,{\tilde X}_{t})\,dt$ is not
constantly equal $6$, we can determine a set of paths $({\tilde{X}}%
_{t}(\omega ))$ for which $\tilde{V}$ is strictly larger than $6$. In view
of Figure 1 it is natural to consider paths which are large, i.e.\ ${%
\tilde{X}}_{t}(\omega )\in \lbrack 8,10]$, for $t\in ]1,2-\tfrac{1}{10}]$
and small, i.e.\ $|{\tilde{X}}_{t}(\omega )|\leq 1$, on the interval $%
]2,3]$. A short \emph{mathematica}-calculation reveals that $\tilde{V}\left(
\omega \right) \gtrsim 6.65>6$ for each such path and according to the
full-support statement the set of all such paths has positive probability,
hence $\tilde{V}$ is indeed not deterministic.

Using elementary analysis it is not difficult to turn numerical
evidence into rigorous mathematics. 
Making \eqref{ExplicitSigma} explicit yields that $\sigma^2_{loc}(t,x)\equiv 2$ for $t\in [0,1]$ and that
\begin{align}
\sigma^2_{loc}(t+1,x)=\frac{\tfrac{3}{\sqrt{2+3t}} e^{-\tfrac{(2x+2+3t)^2}{8(2+3t)}}+\tfrac{1}{\sqrt{2+t}}e^{-\tfrac{(2x+2+t)^2}{8(2+t)}}}{\tfrac{1}{\sqrt{2+3t}} e^{-\tfrac{(2x+2+3t)^2}{8(2+3t)}}+\tfrac{1}{\sqrt{2+t}}e^{-\tfrac{(2x+2+t)^2}{8(2+t)}}}
\end{align}
\begin{align}
\sigma^2_{loc}(t+2,x)=\frac{\tfrac{1}{\sqrt{5+t}} e^{-\tfrac{(2x+5+t)^2}{8(5+t)}} +\tfrac{3}{\sqrt{3+3t}}e^{-\tfrac{(2x+3+3t)^2}{8(3+3t)}}}{\tfrac{1}{\sqrt{5+t}} e^{-\tfrac{(2x+5+t)^2}{8(5+t)}} +\tfrac{1}{\sqrt{3+3t}}e^{-\tfrac{(2x+3+3t)^2}{8(3+3t)}}} 
\end{align}
for $t\in ]0,1]$. We fix  $\varepsilon \in ]0,1]$ and observe that it is simple to see that $\lim_{x\to \infty} \sigma^2_{loc}(t+1,x)= 3$, uniformly w.r.t.\ $t\in [\varepsilon,1]$,  and that $\lim_{x\to 0} \sigma^2_{loc}(t+1,x)\geq 2$, uniformly w.r.t.\ $t\in ]0,1]$. 
It follows that there exists some $\delta>0$ such that 
\begin{align*}
&\sigma^2_{loc}(t+1,x)\geq 3-\varepsilon \ \mbox {for $ x>\tfrac 1\delta, t\in [\varepsilon,1]$ and }\\
&\sigma^2_{loc}(t+1,x)\geq 2-\varepsilon\  \mbox {for $ |x|<\delta, t\in ]0,1]$.}
\end{align*}
Thus  we obtain
\begin{equation}\label{IntegralBound}
\tilde V(\omega)=\int_{0}^{3}{\sigma }_{loc}^{2}(t,{\tilde X}_{t}(\omega))\,dt\geq 1\cdot 2+ (1-2\varepsilon)\cdot (3-\varepsilon)+1\cdot (2-\varepsilon)
\end{equation}
for every path $\tilde X(\omega)$ satisfying $\tilde X_t(\omega)>\tfrac1\delta$ for $t\in [1+\varepsilon, 2-\varepsilon]$ and $|\tilde X_t(\omega)|<\delta$ for $t\in [2,3]$.  
 This set of paths $\tilde X(\omega)$  has positive probability and the quantity on the right side of \eqref{IntegralBound} is strictly larger than $6$ provided that $\varepsilon$ was chosen sufficiently small. Hence we find that $\tilde V$ is not constantly equal to $6$ as required.
  
\medskip

For what it's worth, the example can be modified such that volatility is adapted
to the filtration of the driving Brownian motion.

The trick is to choose a random sign $\hat \epsilon, P(\hat \epsilon=+1)=P(\hat \epsilon=-1)=\tfrac{1}{2}$ depending solely on the behavior of $(B_{t})_{0\leq t\leq 1}$ and in such a way that   $S_1$ is \emph{independent} of $\hat \epsilon $.  
For instance, if we let $m(s)$ be the  unique number satisfying $%
P(S_{1/2}>m(s)|S_{1}=s)=P(S_{1/2}\leq m(s)|S_{1}=s)=\tfrac{1}{2}$, it is sensible to define $\hat\epsilon:=+1$ if $S_{1/2}>m(S_{1})$ and $\hat \epsilon :=-1 $ otherwise. 

We then leave the stock price process unchanged on $[0,1]$, i.e.\ 
we define $\hat{\sigma}^2(t)={\sigma}^2(t)=2$ and $\hat S_t =S_t$ for $t\in [0,1]$. On $]1,2]$ resp.\ $]2,3]$ we set  $\hat \sigma^2(t):=2 +\hat \epsilon$ resp.\ $\hat \sigma^2(t):=2 -\hat \epsilon$
and define $\hat S_t, t\in ]1,3]$ as the solution  of the SDE 
\begin{align}\label{SameSDE}
d\hat S_t=\hat \sigma(t) \hat S_t\, dB_t,\ \hat S_1= S_1  .
\end{align}
Here \eqref{SameSDE} depends only on $S_1$ and the process $(B_t-B_1)_{1\leq t\leq 3}$; since both are independent of $\hat \epsilon$, we obtain that $(\hat S_t)_{1\leq t\leq 3}$ and $(S_t)_{1\leq t\leq 3}$ are equivalent in law. 
 It follows that $ \hat V=\int_{0}^{3}\hat\sigma ^{2}(t,\omega )\,dt\equiv 6 $ and since $\hat S_t$ and $ S_t$ have the same law for each $t\in [0,3]$, they induce the same local volatility model and in particular the same (non deterministic) $\tilde V$.


\section{Counterexample for a Markovian stochastic volatility model}
Recall that $X$ denotes the log-price process of a general stochastic
volatility model; 
\begin{equation*}
dX_{t}=\sigma \left( t,\omega \right) dB_{t}-\left( \sigma ^{2}\left(
t,\omega \right) /2\right) dt,
\end{equation*}%
where $\sigma =\sigma \left( t,\omega \right) $ is the (progressively
measurable) instantaneous volatility process\footnote{%
We note that we intend to insert later the volatility of example \ref{toyexample}, but so far our considerations hold in general.}. Recall also our standing
assumption that $\sigma $ is bounded from above and below by positive
constants. We would like to apply theorem \ref{ThmGy} to the 2D diffusion $%
\left( X,V\right) $ where $dV=\sigma ^{2}dt$ keeps track of the running
realized variance\footnote{%
In other words, 
$V_{T}=\int_{0}^{T}\sigma ^{2}\left( t,\omega \right) dt,$
if $V_{0}=0$ which we shall assume from here on.}. We can only do so
after elliptic regularization. That is, we consider%
\begin{eqnarray*}
dX_{t} &=&\sigma \left( t,\omega \right) dB_{t}-\left( \sigma ^{2}\left(
t,\omega \right) /2\right) dt, \\
da_{t}^{\varepsilon } &=&\sigma ^{2}\left( t,\omega \right) dt+\varepsilon
^{1/2}dZ_{t}
\end{eqnarray*}%
where $Z$ is a\ Brownian motion, independent of of the filtration generated by $B$ and $\sigma$. It follows that the following ``double-local'' volatility model%
\begin{eqnarray*}
d\tilde{X}_{t}^{\varepsilon } &=&\sigma _{dloc}\left( t,\tilde{X}%
_{t}^{\varepsilon },\tilde{a}_{t}^{\varepsilon }\right) dB_{t}-\left( \sigma
_{dloc}^{2}\left( t,\tilde{X}_{t}^{\varepsilon },\tilde{a}_{t}^{\varepsilon
}\right) /2\right) dt, \\
d\tilde{a}_{t}^{\varepsilon } &=&\sigma _{dloc}^{2}\left( t,\tilde{X}%
_{t}^{\varepsilon },\tilde{a}_{t}^{\varepsilon }\right) dt+\varepsilon
^{1/2}dZ_{t}, 
\end{eqnarray*}%
(with $\sigma _{dloc}^{2}\left( t,x,a\right) =E[ \sigma
^{2}\left( t,\omega \right) |X_{t}=x,a_{t}^{\varepsilon }=a])$
has the one-dimensional marginals of the original process $\left(
X_{t},a_{t}^{\varepsilon }\right) $. That is, for all fixed $t$ and $%
\varepsilon $,%
\begin{equation*}
X_{t}\overset{law}{=}\tilde{X}_{t}^{\varepsilon }\text{ and }\tilde{a}%
_{t}^{\varepsilon }\overset{law}{=}a_{t}^{\varepsilon }\text{.}
\end{equation*}%
Let us also note that the law of $a_{t}^{\varepsilon }$ is the law of $%
V_{t}=a_{t}^{0}$ convolved with a standard Gaussian of mean $0$ and
variance $\varepsilon $. 
The log-price processes $X$ and $\tilde{X}^{\varepsilon }$ induce the
same local volatility surface. To this end, just observe that $X_{t}\overset%
{law}{=}\tilde{X}_{t}^{\varepsilon }$ implies identical call option prices
for all strikes and maturities and hence (by Dupire's formula)\ the same
local volatility:%
\begin{equation*}
\sigma _{loc}^{2}\left( t,x\right) =E\big[ \sigma ^{2}\big( t,\omega
\big) |X_{t}=x\big] =E\big[ \sigma _{dloc}^{2}\big( t,\tilde{X}%
_{t}^{\varepsilon },\tilde{a}_{t}^{\varepsilon }\big) |\tilde{X}%
_{t}^{\varepsilon }=x\big] .
\end{equation*}%
Since the law of a time inhomogeneous Markov process is fully specified by its generator, 
it follows that the law of the local volatility process associated to $(X)$ has the same
law as the local volatility process associated to $( \tilde{X}^{\varepsilon })$. 

We apply this to the toy model discussed earlier. Recall that in
this example, with $T=3$%
\begin{equation*}
V_{T}=\int_{0}^{T}\sigma ^{2}\left( t,\omega \right) dt=6
\end{equation*}%
whereas realized variance under the corresponding local volatilty model,%
\begin{equation*}
\tilde{V}_{T}=\int_{0}^{T}\sigma^{2}_{loc}\left( t,\tilde{X}_{t}\right) dt
\end{equation*}%
was seen to be random (but with mean $V_{T}$, thanks to the matching
variance swap prices). As a particular consequence, using Jensen%
\begin{eqnarray*}
E\Big(\int_{0}^{T}\sigma _{loc}^{2}\big( t,\tilde{X}_{t}\big) dt-6\Big)^+ &>&
 \Big( E \int_{0}^{T}\sigma _{loc}^{2}\big( t,\tilde{X}_{t}\big) dt-6\Big)^+\\
&=& \Big( E \int_{0}^{T}\sigma ^{2}\left( t,\omega \right) dt-6\Big)^+
=(V_{T}-6)^+ = 0.
\end{eqnarray*}%
We claim that this persists when replacing the abstract stochastic
volatility model $( X) $ by $( \tilde{X}^{\varepsilon
}) $, the first component of a 2D Markov diffusion, for any $\varepsilon
>0 $. Indeed, thanks to the identical laws of the respective local volatility processes the
left-hand side above does not change when replacing $(\tilde{X})$  by the local volatility process associated to $( \tilde{X}^{\varepsilon })$.  On the other hand%
\begin{eqnarray*}
E\int_{0}^{T}\sigma _{dloc}^{2}\big( t,\tilde{X}_{t}^{\varepsilon },\tilde{a%
}_{t}^{\varepsilon }\big) dt &=&E(\tilde{a}_{T}^{\varepsilon }-\varepsilon
^{1/2}Z_{T}) \\
&=&E\left( \tilde{a}_{T}^{\varepsilon }\right) =E\left( a_{T}^{\varepsilon
}\right) \\
&=&E\big( V_{T}+\varepsilon ^{1/2}Z_{T}\big) =V_{T}.
\end{eqnarray*}%
Thus, insisting again that the process $\tilde{X}$ is (in law) the local
volatility model associated to the double local volatility model $( \tilde{X}^{\varepsilon }, \tilde{a}^{\varepsilon}) $ we
see that
\begin{equation*}
c=E\Big(\int_{0}^{T}\sigma _{loc}^{2}\big( t,\tilde{X}_{t}\big) \, dt - 6\Big)^+>
\Big( E\int_{0}^{T}\sigma _{dloc}^{2}\big( t,\tilde{X}_{t}^{\varepsilon },\tilde{a%
}_{t}^{\varepsilon }\big) dt-6\Big)^+ = 0.
\end{equation*}
(Observe that $c>0$ is independent of $\varepsilon$.) Using the Lipschitz property of the hockeystick function, again Gy\"{o}ngy and the fact that $a^\varepsilon_T$ is normally distributed with mean $V_T$ and variance $\varepsilon T$ we can conclude that
\begin{eqnarray*}
E\Big( \int_{0}^{T}\sigma _{dloc}^{2}\big( t,\tilde{X}_{t}^{\varepsilon },\tilde{a%
}_{t}^{\varepsilon }\big) dt-6\Big)^+\!\!\! &\!\!=\!\! &\!E\Big( \int_{0}^{T}\sigma _{dloc}^{2}\big( t,\tilde{X}_{t}^{\varepsilon },\tilde{a%
}_{t}^{\varepsilon }\big) dt+ \varepsilon^\frac{1}{2} Z_T - 6  -\varepsilon^\frac{1}{2} Z_T \Big)^+ \\ 
&\!\! \leq\!\! &\!  E\big((\tilde{a}_{T}^{\varepsilon }-6)^+ + \vert \varepsilon^\frac{1}{2}Z_T \vert\big) \\
&\!\!=\!\! &\! E(a_{T}^{\varepsilon }-6)^+ + E\vert \varepsilon^\frac{1}{2}Z_T \vert
= 3\sqrt{\varepsilon T/2\pi}.
\end{eqnarray*}
Now we choose $\varepsilon$ small enough such that $3 \sqrt{\varepsilon T/2\pi}<c$, whence the conjecture fails to hold true in the double local volatility model for $\varepsilon >0$ small enough.
\section{Conclusion}
Summing up, the double-local volatility model constitutes an example of a continuous
2D Markovian stochastic volatility model, where stochastic volatility is a
function of both state variables,  in which conjecture \ref{ConjDup} fails,
i.e.\ in which the minimal possible value of a call option is not generated by a local volatility model.



\end{document}